\documentclass[aps,showpacs,a4paper,floatfix,twocolumn,prc,amsmath,amssymb]{revtex4-1}
\usepackage{amsmath}
\usepackage{graphicx}
\usepackage{ulem}
\usepackage{eucal}
\usepackage{morefloats}
\usepackage{rotating}
\usepackage{relsize}
\usepackage{color}
\usepackage{floatflt,epsfig}
\usepackage{ulem}
\usepackage{natbib}
\usepackage{hyperref}

\begin{document}

\title{ Neutron stars in the context of $f(\mathbb{T},\CMcal{T})$ gravity}

\author{Cl\'esio E. Mota$^1$}
\email{clesio200915@hotmail.com}
\author{Luis C. N. Santos$^2$}
\email{luis.santos@ufsc.br}
\author{Franciele M. da Silva$^3$}
\email{franmdasilva@gmail.com}
\author{Cesar V. Flores$^{4,5}$}
\email{cesarovfsky@gmail.com}
\author{Iarley P. Lobo$^{6,7}$}
\email{iarley_lobo@fisica.ufpb.com}
\author{Valdir B. Bezerra$^2$}
\email{valdirbarbosa.bezerra@gmail.com}

\affiliation{$^1$Departamento de F\'isica, CFM - Universidade Federal de Santa Catarina; C.P. 476, CEP 88.040-900, Florian\'opolis, SC, Brasil.}

\affiliation{$^2$Departamento de F\'isica, CCEN-Universidade Federal da Para\'iba; \\ C.P. 5008, CEP  58.051-970, João Pessoa, PB, Brazil}

\affiliation{$^3$N\'ucleo Cosmo--ufes \& Departamento de F\'isica, Universidade Federal do Esp\'irito Santo, Av. Fernando Ferrari, 540, CEP 29.075-910, Vit\'oria, ES, Brazil}

\affiliation{$^4$Centro de Ci\^encias Exatas, Naturais e Tecnol\'ogicas, CCENT - Universidade Estadual da Regi\~ao Tocantina do Maranh\~ao; C.P. 1300,\\ CEP 65901-480, Imperatriz, MA, Brasil.}

\affiliation{$^5$Departamento de F\'isica, CCET - Universidade Federal do Maranh\~ao, Campus Universit\'ario do Bacanga; CEP 65080-805, S\~ao Lu\'is, MA, Brasil.}

\affiliation{$^6$Department of Chemistry and Physics, Federal University of Para\'iba, Rodovia BR 079 - Km 12, 58397-000 Areia-PB,  Brazil.}

\affiliation{$^7$Physics Department, Federal University of Lavras, Caixa Postal 3037, 37200-000 Lavras-MG, Brazil.}

\begin{abstract}
In this work, we investigate the existence of neutron stars (NS) in the framework of $f(\mathbb{T},\CMcal{T})$ gravity, where $\mathbb{T}$ is the torsion tensor and $\CMcal{T}$ is the trace of the energy-momentum tensor. The hydrostatic equilibrium equations are obtained, however, with $p$ and $\rho$ quantities passed on by effective quantities $\Bar{p}$ and $\Bar{\rho}$, whose mass-radius diagrams are obtained using modern equations of state (EoS) of nuclear matter derived from relativistic mean field models and compared with the ones computed by the Tolman-Oppenheimer-Volkoff (TOV) equations. Substantial changes in the mass-radius profiles of NS are obtained even for small changes in the free parameter of this modified theory. The results indicate that the use of $f(\mathbb{T},\CMcal{T})$ gravity in the study of NS provides good results for the masses and radii of some important astrophysical objects, as for example, the low-mass X-ray binary (LMXB) NGC 6397 and the pulsar of millisecond PSR J0740+6620. In addition, radii results inferred from the Lead Radius EXperiment (PREX-2) can also be described for certain parameter values.

\vspace*{0.5cm}

\end{abstract}
\maketitle

\section{Introduction}

In recent years, there have been a growing number of ideas exploring modifications and alternative formulations of General Relativity (GR) emerging of different contexts. In fact, GR is a theory well tested, providing an interesting description of the space-time nature as a dynamical stage where physical phenomena takes place. In parallel to the advances in GR, the quantization of the gravitational field remains an open problem. With respect to this issue, it was pointed out that the action for gravity should be constructed with higher-order curvature terms in the context of renormalization at one loop level \cite{witt}. In the literature there are some formulations of gravity where the usual Einstein-Hilbert action is supplemented by higher-order curvature terms, as for example in the context of the $f(R)$ theory in which case the Ricci scalar $R$ in the action is replaced by a general function  $f(R)$ \cite{fder1}. 

On the other hand, there are questions concerning the content of energy and matter in the universe that, at the moment, are not satisfactorily explained in the scope of standard theories. The observed rotation curves of galaxies \cite{vera} and the ``missing mass'' of galaxy clusters \cite{cluster} suggest the dark matter hypothesis, while the accelerated expansion of the universe observed today can be interpreted as an effect of the so-called dark energy \cite{Perlmutter:1998np,Sahni:2004ai}. Unexpectedly these observations reveals that the ordinary baryonic matter corresponds to only $4\%$ of content of energy of the universe while the dark matter and dark energy correspond to $20\%$ and $76\%$, respectively. In this sense, there are studies considering the possibility of modified theories of gravity which may help to alleviate the need for dark components of energy of the universe beyond the scope of GR.

The late-time acceleration of the universe can be interpreted under two points of view. In the first one, it is introduced a dark energy sector in the energy content of the universe through a type of field. In the second one, the gravitational field itself is modified. In addition, there may be combinations of both approaches depending on the couplings between gravitational and non-gravitational sectors of theory \cite{capozziello2011extended,de2010f,nojiri2011unified,lobo2008dark}. Thus, it is expected that different formulations of gravity imply that standard results in astrophysics suffer modifications. Compact objects as neutron stars (NS), have been studied considering effects of such modifications \cite{harada1998neutron,orellana2013structure,momeni2015tolman,oliveira2015neutron,hendi2016modified,singh2019einstein,maurya2020charged,mota2019combined,paper1,da2021rapidly}. NS in the context of $f(R)$ gravity were studied in \cite{ns1,ns2,ns3} and in $f(R,T)$ gravity in the papers \cite{ns4,ns5,ns6,ns7,ns8}. In common, all of these works have considered effects on NS due to the modification of the gravitational field that include extra terms in the action. In the scheme of nonconservative gravity, the modification of the gravitational field can be done through a reinterpretation of the conservation law, as was considered in the papers \cite{rastall,santos2022} (for a review
on non-conservative theories of gravity, see \cite{Velten:2021xxw}). Usually, the non-conservation of the stress-energy tensor is proportional to the matter density and pressure themselves. For this reason, an environment such as a compact object like a NS turns out to be an appealing laboratory for testing such theories.

In the context of modified theories of gravity, the so-called $f(\mathbb{T}$,$\CMcal{T})$ gravity is a class of such theories, free of ghosts and instabilities which, when applied to cosmological problems, leads to interesting results \cite{harko2014f}. In this formulation, the action depends on the torsion scalar $\mathbb{T}$ and on the trace of the energy-momentum tensor $\CMcal{T}$. As in the case of $f(\mathbb{T})$ gravity where the action is an arbitrary function of the torsion, in $f(\mathbb{T},\CMcal{T})$ gravity, the action is a arbitrary function of both the trace of the energy-momentum tensor and the torsion scalar.

In this paper, we study an important context, not yet explored in the literature, that are the implications of the $f(\mathbb{T}$,$\CMcal{T})$ gravity on NS. In particular, we obtain the mass-radius relation of NS in the context of this modified gravity and compare our results with recent astrophysical observations and experiments. 

This work is organized as follows: In Section \ref{sec1} we expose a summary of the $f(\mathbb{T}$,$\CMcal{T})$ gravity. In Section \ref{sec2} we derive the equations describing static, spherically symmetric stars in this modified theory of gravity. In Section \ref{sec3} we present our results and in Section \ref{sec4}
we close with our final remarks.

\section{Gravitational Field Equations of $f(\mathbb{T},\CMcal{T})$ Gravity} \label{sec1}

Given a line element describing a space-time we want to study
\begin{equation}
ds^2=g_{\mu\nu}dx^\mu dx^\nu=\eta_{AB}e^{A}_{\;\;\mu}e^{B}_{\;\;\nu}dx^{\mu}dx^{\nu}
\label{eq1}
\end{equation}
where $g_{\mu\nu}$ and $\{e^{A}_{\;\mu}\}$ are respectively the metric tensor and the components of the tetrad associated to space-time geometry, and $\eta_{AB}=diag(1,-1,-1,-1)$ is the Minkowski metric. The signature $(+\;-\;-\;-)$ and geometrized units, that is, $G = c = 1$, will be taken into account. In GR we assume that gravity is associated with the curvature of the space-time and thus we use the Levi-Civita's connection
\begin{equation}
\overset{\circ}{\Gamma }{}_{\;\;\mu \nu }^{\rho } =
\frac{1}{2}g^{\rho \sigma }\left(
\partial _{\nu} g_{\sigma \mu}+\partial _{\mu}g_{\sigma \nu}-\partial _{\sigma}g_{\mu \nu}\right)
\label{eq2}
\end{equation}
to compute quantities associated with the curvature such as the Ricci scalar, $R$, that is present in the GR's action.

On the other hand, in teleparallel theory one assumes that gravity is associated to the torsion of the space-time and thus the Weizenbock's connection
\begin{equation}
    \Gamma^{\lambda}_{\mu\nu}=e^{\;\;\lambda}_{A}\partial_{\mu}e^{A}_{\;\;\nu}=-e^{A}_{\;\;\mu}\partial_\nu e_{A}^{\;\;\lambda}
    \label{eq3}
\end{equation}
is used to construct quantities associated with the torsion, as the torsion scalar $\mathbb{T}$ that appears in the teleparallel gravity action. In the modified teleparallel theories it is assumed that the action depends on a arbitrary function of $\mathbb{T}$. In our case, we are going to consider a modified action given by \cite{harko2014f}
\begin{equation}
\mathbb{S}= \int d^{4}x~~e \left[\frac{\mathbb{T}+f(\mathbb{T},\CMcal{T})}{16\pi}+\mathcal{L}_{m}\right],
\label{eq4}
\end{equation}
where $e$ is the determinant of the tetrads $e=\text{det}(e^A_{\;\;\mu})=\sqrt{-g}$ and $\CMcal{T}=g^{\mu\nu}T_{\mu\nu}$ is the trace of the energy-momentum tensor $T_{\mu\nu}$, which can be obtained from the Lagrangian for the matter distribution ${\mathcal{L}}_m$ in the following way 
\begin{equation}
    T_{\mu\nu}=g_{\mu\nu} {\mathcal{L}}_m - 2 \frac{\partial {\mathcal{L}}_m}{\partial {g^{\mu\nu}}}\, .
    \label{eq5}
\end{equation}
Let us assume that the function $f(\mathbb{T},\CMcal{T})$ is given by
\begin{equation}
    f\left(\mathbb{T},\CMcal{T}\right)=\overline{\omega}\, \mathbb{T}^n\,
\CMcal{T} -2  \Lambda\, ,
\label{eq6}
\end{equation}
where $\overline{\omega}$, $n$ and $\Lambda $ are arbitrary constants, specifically $\overline{\omega}$ can be interpreted as a coupling constant of geometry with matter fields, $n$ is a pure number (assumed to be unity here) and $\Lambda $ can be recognized as the cosmological constant as discussed in \cite{harko2014f,salako2020study}.  

We are interested in matter that can be described by a perfect fluid, so that $T_{\mu\nu}$ is given by:
\begin{equation}
    T_{\mu\nu} = - p g_{\mu\nu} + ( p + \rho ) u_{\mu} u_{\nu},
    \label{eq7}
\end{equation}
where $p$ is the pressure and $\rho$ is the energy density of the fluid. By varying the action from Equation (\ref{eq4}) with respect to the tetrad we find the following field equation
\begin{equation}
    G_{\mu\nu} = 8\pi T_{\mu\nu}^{eff},
    \label{eq8}
\end{equation}
where the effective energy-momentum tensor $T_{\mu\nu}^{eff}$ is 
\begin{equation}
    T_{\mu\nu}^{eff} = g_{\mu\nu}\left[\frac{\Big(-\overline{\omega}(\rho - 3p) + 2\Lambda\Big)}{16\pi} + \frac{\overline{\omega}p}{8\pi}\right] + T_{\mu\nu}\bigg(1 + \frac{\overline{\omega}}{8\pi}\bigg).
    \label{eq9}
\end{equation}
Calculating the covariant derivative of the  energy-momentum tensor given by Equation (\ref{eq7}), we obtain the following result
\begin{equation}
    \nabla_{\mu}T_{\nu}{}^{\mu} = \frac{1}{\Big(4\pi + (1/2)\overline{\omega}\Big)}\left\{\frac{\overline{\omega}}{4}(\partial_{\nu}\CMcal{T}) - \frac{\overline{\omega}}{2}\partial_{\nu}p \right\}.
    \label{conservation}
\end{equation}
In a cosmological context, equation $\ref{conservation}$  can be associated to creation or destruction of matter throughout the universe evolution. As discussed in \cite{ns6}, the interpretation of  creation or destruction of matter particles in the NS level encounters difficulties in a static framework as occurs in the study of the hydrostatic equilibrium expression, i.e, the Tolman-Oppenheimer-Volkof equation. Also, it usually implies in the presence of a fifth force and non-geodesic trajectory for free particles. Naturally, results that depend on such imput would also be modified correspondingly. However, this is not the case analyzed in the present paper.
In the next section we use Equations (\ref{eq8}) to (\ref{conservation}) to obtain and analyse the mass-radius relation of NS in the context of modified teleparallel gravity.

\section{Stellar Structure Equations} \label{sec2}

In this section, we discuss some of the main procedures that leads to the deduction of the hydrostatic equilibrium equation in the context of $f(\mathbb{T},\CMcal{T})$ gravity. 

To study compact stars, such as NS, magnetars and other astrophysical structures, we assume these objects as being homogeneous, static (no rotation), isotropic and spherically symmetric \cite{Carroll2004}. Therefore, we must use the appropriate metric in a convenient coordinate system that describes the object being studied. The most general metric describing the space-time under consideration is given by the line element
\begin{equation}
    ds^{2}=e^{\nu(r)}dt^{2}-e^{\lambda(r)}dr^{2}-r^{2}(d\theta^{2}+\sin{\theta}^{2}d\phi^{2}),
    \label{eq10}
\end{equation}
where $\nu$ and $\lambda$ are radial functions that we want to determine based on the field equations (\ref{eq8}). Thus, using Equation (\ref{eq10}) and substituting appropriately into Equation (\ref{eq8}),we obtain the following results
\begin{widetext}
\begin{equation}
    e^{-\lambda}\Big(\frac{\lambda'}{r} - \frac{1}{r^2}\Big) + \frac{1}{r^2} = 8\pi\left\{\left[ \frac{\Big(-\overline{\omega}(\rho - 3p) + 2\Lambda\Big)}{16\pi} + \frac{\overline{\omega} p}{8\pi}\right] + \rho \Bigg(1 + \frac{\overline{\omega}}{8\pi}\Bigg)\right\} = 8\pi\Bar{\rho},
    \label{eq11}
\end{equation}
\begin{equation}
    e^{-\lambda}\Big(\frac{\nu'}{r} + \frac{1}{r^2}\Big) - \frac{1}{r^2} = - 8\pi\left\{\left[ \frac{\Big(-\overline{\omega}(\rho - 3p) + 2\Lambda\Big)}{16\pi} + \frac{\overline{\omega} p}{8\pi}\right] - p \Bigg(1 + \frac{\overline{\omega}}{8\pi}\Bigg)\right\} = 8\pi \Bar{p},
    \label{eq12}
\end{equation}
\begin{equation}
  \frac{e^{-\lambda}}{4r}\left[2\Big(\lambda'-\nu'\Big) - \Big(2\nu''+\nu'^{2} - \nu'\lambda'\Big)r\right] = - 8\pi\left\{\left[ \frac{\Big(-\overline{\omega}(\rho - 3p) + 2\Lambda\Big)}{16\pi} + \frac{\overline{\omega} p}{8\pi}\right] - p \Bigg(1 + \frac{\overline{\omega}}{8\pi}\Bigg)\right\} = 8\pi \Bar{p},
  \label{eq12_1}
\end{equation}
\end{widetext}
where, the prime denotes a derivative with respect to the radial coordinate $r$. The quantities $\Bar{\rho}$ and $\Bar{p}$ are the effective pressure and energy density, defined as
\begin{equation}
    \Bar{\rho} = \rho + \frac{\overline{\omega}\rho}{16\pi} + \frac{5\overline{\omega} \ p}{16\pi} + \frac{\Lambda}{8\pi}, 
    \label{eq13}
\end{equation}
\begin{equation}
    \Bar{p} = p + \frac{\overline{\omega}\rho}{16\pi} - \frac{3\overline{\omega} \ p}{16\pi} - \frac{\Lambda}{8\pi}. 
    \label{eq14}
\end{equation}
In addition to the field equations, we also need to consider the conservation equation (\ref{conservation}) in $f(\mathbb{T},\CMcal{T})$ gravity so that we have a complete set of equations to be solved. In the case we are studying, Equation (\ref{conservation}) has the form as follows
\begin{equation}
    -p' - \frac{\nu'}{2}(\rho + p) = \frac{1}{\Big(4\pi + (1/2)\overline{\omega}\Big)}\left\{ \frac{\overline{\omega}\rho'}{4} - \frac{5\overline{\omega} \ p'}{4}\right\}.
    \label{eq15}
\end{equation}
Redefining the function $\lambda(r)$ as 
\begin{equation}
    e^{-\lambda(r)} = 1 - \frac{2 M(r)}{r},
    \label{eq16}
\end{equation}
and rearranging Equations (\ref{eq11}) and and (\ref{eq15}), we get the equations required to describe static spherically symmetric stellar structures in $f(\mathbb{T},\CMcal{T})$ gravity theory, which are given by
\begin{equation}
\frac{d M}{dr}=4 \pi r^{2} \Bar{\rho},
\label{eq17}
\end{equation}
and
\begin{equation}
\frac{d\Bar{p}}{dr} =-\frac{M\Bar{\rho}}{r^{2}}\left[1+\frac{\Bar{p}}{\Bar{\rho}}\right] \left[1+\frac{4\pi r^{3} \Bar{p}}{M} \right]\left[1-\frac{2M}{r}\right]^{-1}.
\label{eq18}
\end{equation}  
In the next section we show some results obtained by solving Equations (\ref{eq17}) and (\ref{eq18}) for realistic EoS of NS.

\section{Results} \label{sec3}

In this section, we present the results obtained from the solution of the field equations in the context of $f(\mathbb{T},\CMcal{T})$ modified theory of gravity applied to NS.

As an input to the stellar hydrostatic equilibrium equations, we use two realistic EoS obtained from a relativistic mean field (RMF) approach. Firstly, we consider the IU-FSU \cite{PhysRevC.82.055803} parametrization because it is able to explain reasonably well both nuclear \cite{PhysRevC.99.045202} and stellar matter properties \cite{PhysRevC.93.025806}. We then compare the IU-FSU results with the ones obtained with a stiffer EoS calculated with a model of coupling of mesons and quarks, the quark–meson coupling (QMC) model \cite{guichon}. (For the EoS with the QMC model, we refer the reader to refs. \cite{guichon,saito94,saito95,Pal95,Grams:2015inf}.) It is well known that a stiffer EoS leads to a bigger NS maximum mass in contrast to a softer one. In fact, using the EoS QMC as an input to the stellar equilibrium equations yields a maximum mass greater than 2.0 M$_\odot$, and, therefore, we want to verify that we get the same qualitative behavior for macroscopic properties (such as mass and radius) with parameterizations that are substantially different. For the NS crust, we use the full BPS \cite{bps} EoS.

After defining the EoS, some boundary conditions are required to solve the equations (\ref{eq17}) and (\ref{eq18}) along the radial coordinate $r$, from the center towards the surface of the star. At the star’s center $r=0$ we take:
\begin{equation}
M (0) = 0 \ ; \ \ \Bar{\rho} (0) = \Bar{\rho}_{c} \ ; \ \ \Bar{p}(0) = \Bar{p}_{c}.    
\end{equation}
The radius of the star $(r = R)$ is determined as the point where the pressure vanishes, \textit{i.e}, $\Bar{p}(R) = 0$. At this point, the interior solution connects softly with the Schwarzschild vacuum solution, indicating that the potential metrics of the interior and the exterior metric are
related as $e^{\nu(R)} = \frac{1}{e^{\lambda(R)}} = 1 - 2M/R$, being $M$ the total mass of the star.

Let us discuss and compare our results with recent astrophysical observations and nuclear physics experiments. At first, the NS in LMXB NGC 6397, depicted as a green shaded area in all figures, provides a constraint at 68\% confidence level over the possible values of the masses and corresponding radii of the NS \cite{Ozel_2016, Steiner_2018}. Similarly, the millisecond pulsars are among the most useful astrophysical objects in the Universe for testing fundamental physics, because they impose some of the most stringent constraints on high-density nuclear physics in the stellar interior \cite{Cromartie}. Recent measurements coming from the \textit{Neutron Star Interior Composition Explorer} (NICER) mission reported pulsar observations for canonical ($1.4~$M$_\odot$) and massive ($2.0~$M$_\odot$) NS. The mass measurement and radius estimates provided for these objects, are 11.80 km $\leq R_{1.4} \leq$ 13.1 km for the $1.4 M_\odot$ NS PSR J0030+0451 (horizontal line segment in red colour shown in all Figures) and 11.60 km $\leq R \leq$ 13.1 km for a NS with mass between $2.01  M_\odot \leq M \leq 2.15  M_\odot$ PSR J0740+6620 (the rectangular region in orange colour shown in all Figures). However, the authors of Ref. \cite{PREX2021} used the recent measurement of neutron skin on $^{208}$Pb by PREX-2 to constrain the radius of NS, which leads to a prediction of the radius of the canonical $1.4~M_\odot$ of 13.25 km $\lesssim R_{1.4} \lesssim$ 14.26 km (horizontal line segment in green colour shown in all Figures). Likewise, we also compare our results with two massive stars that had been discovered in 2010 and 2013, namely, PSR J1614+2230 \cite{Demorest} with mass $1.97 \pm 0.04$ $M_\odot$ (horizontal line in blue colour shown in all Figures) and PSR J0348+0432 \cite{Antoniadis} with mass $2.01 \pm 0.04$ $M_\odot$ (horizontal line in pink colour shown in all Figures). Our results are discussed in the next paragraphs.

\begin{figure}[t]
    \centering
    \includegraphics[width=6.5cm,angle=270]{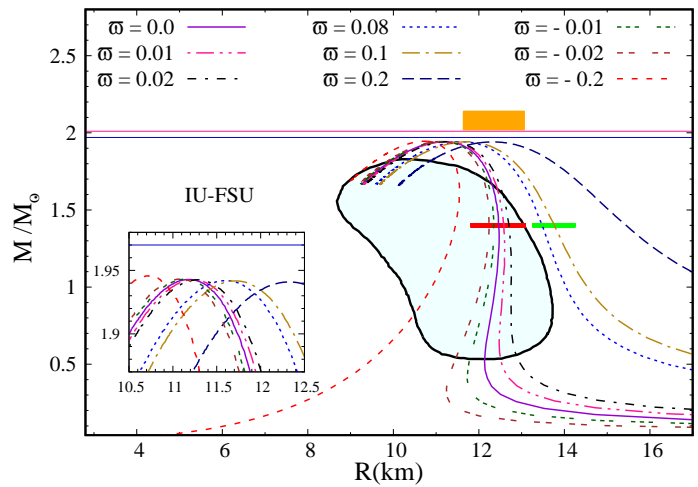}
    \caption{Mass-radius relation for families of NS's described by the IU-FSU EoS. We analyse the effect of varying the parameter $\overline{\omega}$ of the $f(\mathbb{T},\CMcal{T})$ theory. The red and green line segment represent the radius range of the $1.4 M_\odot$ NS for PSR J0030 + 0451 and PREX-2, respectively. The orange rectangular region corresponds to the range of radius estimates for $2.08 \pm 0.07 M_\odot$ NS PSR J0740+6620. Similarly, the blue and pink horizontal lines stand, respectively, for the mass measurements of NS PSR J1614 + 2230 and NS PSR J0348 + 0432. The purple solid line curve is solution for the usual TOV equation from GR.}
    \label{fig1}
\end{figure}

\begin{figure}[t]
    \centering
    \includegraphics[width=6.5cm,angle=270]{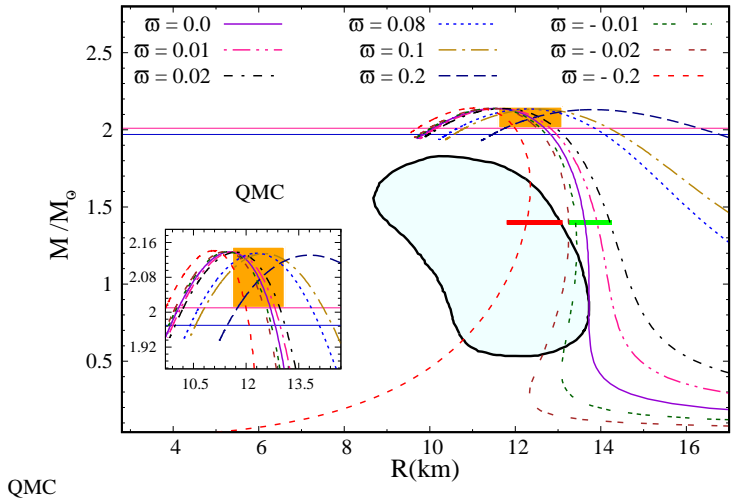}
    \caption{Mass-radius relation for families of NS's described by the QMC EoS. We analyse the effect of varying the parameter $\overline{\omega}$ of the $f(\mathbb{T},\CMcal{T})$ theory. The red and green line segment represent the radius range of the $1.4 M_\odot$ NS for PSR J0030 + 0451 and PREX-2, respectively. The orange rectangular region corresponds to the range of radius estimates for $2.08 \pm 0.07 M_\odot$ NS PSR J0740+6620. Similarly, the blue and pink horizontal lines stand, respectively, for the mass measurements of NS PSR J1614 + 2230 and NS PSR J0348 + 0432. The purple solid line curve is the solution for the usual TOV equation from GR.}
    \label{fig2}
\end{figure}

We modelled the function $f(\mathbb{T},\CMcal{T})$ according to equation (\ref{eq6}). This function model has already been used in recent works as, for example, in \cite{harko2014f,salako2020study}. We explore the values of the parameter $\overline{\omega}$ which range from $-0.2$ to $0.2$. On the other hand, we check that the $\Lambda$ parameter has no significant effect on the mass-radius profiles of NS, since it appears as a constant in the $f(\mathbb{T},\CMcal{T})$ function that we have chosen. Therefore, we use $\Lambda = 0$. Note that we recover the GR solution from $f(\mathbb{T},\CMcal{T})$ theory by assuming that $\overline{\omega}=\Lambda = 0$. These plots are represented by the continuous purple lines in the Figures.

In Figure \ref{fig1} we show the effects of $f(\mathbb{T},\CMcal{T})$ theory on NS properties obtained with the IU-FSU EoS. We can see that the value of $\overline{\omega}$ has a very small influence on the maximum mass of the stars. The radius of the canonical NS ($M = 1.4 M_{\odot} $) is considerably affected. Note a bigger (smaller) radius for the most positive (negative) values of $\overline{\omega}$. We can observe that the results of PREX-2 cannot be described with IU-FSU EoS in the GR, but in $f(\mathbb{T},\CMcal{T})$ theory the solutions with $\overline{\omega}=0.08$ and $\overline{\omega}=0.1$ produce mass and radius that agree with this constraint. However, the solutions obtained with IU-FSU EoS cannot describe the mass and radius of PSR J0740+6620, PSR J1614+2230 and NS PSR J0348+0432 neither on GR nor on $f(\mathbb{T},\CMcal{T})$ theory.

In Figure \ref{fig2} we show the mass-radius relation obtained for QMC EoS in $f(\mathbb{T},\CMcal{T})$ gravity. Again, the effect of the parameter $\overline{\omega}$ is to increase the radius when its values increase positively and to decrease the radius when its values increase negatively. At the same time, the maximum mass changes very little with the variation of $\overline{\omega}$. We can also see that the solutions obtained with the QMC EoS in $f(\mathbb{T},\CMcal{T})$ can accommodate almost all the constraints we are taking into consideration, and with a smaller radius than in GR, if we take $\overline{\omega}=-0.01$ or $\overline{\omega}=-0.02$. The exception is NS PSR J0030+0451 which only can be described with QMC EoS in $f(\mathbb{T},\CMcal{T})$ gravity if we take $\overline{\omega}=-0.2$. We can note that for both EoS analysed we could not find a configuration that satisfies all the constraints at the same time.

We can see that for both EoS's the value of $\overline{\omega}$ has a very small influence on the maximum mass of the stars, on the other hand, the value of the radius of the star with maximum mass increases when we increase the value of $\overline{\omega}$ and decreases when $\overline{\omega}$ decreases. Also for both EoS's, the case $\overline{\omega}=-0.2$ produces mass-radius curves that are typical of quark stars. 

\section{Final remarks} \label{sec4}

We have investigated the effects of $f(\mathbb{T},\CMcal{T})$ gravity on NS assuming these compact objects as being homogeneous, static and isotropic. In this way, we have considered a spherically symmetric space-time and solved the field equations and the hydrostatic equilibrium equation in the context of this modified theory of gravity. This type of system can be transformed into a system with effective pressure and energy density which permitted that the hydrostatic equilibrium equation was obtained through known techniques. For the choice of the $f(\mathbb{T},\CMcal{T})$ function used here, we obtained that this theory can predict NS with almost the same mass and smaller radius than in GR, for a given EoS, that is an interesting result in view of the recent observations. Considering the low-mass X-ray binary (LMXB) NGC 6397 and the pulsar of millisecond PSR J0740+6620, the results obtained using the modified hydrostatic equilibrium equations present good agreement with the observed masses and radii. 

We particularize $f(\mathbb{T},\CMcal{T})$ gravity according to equation (\ref{eq6}). The good results obtained in comparison to GR suggest future extensions of this work, as for example, by taking into consideration different choices of the $f(\mathbb{T},\CMcal{T})$ function, which should be done in a near future. It can be interesting to test, for example, high powers in $\mathbb{T}$ besides and new couplings between $\mathbb{T}$ and $\CMcal{T}$. In addition, we can use different EoS as input to the stellar hydrostatic equilibrium equations along the aforementioned choices of $f(\mathbb{T},\CMcal{T})$ function. 

\section*{Acknowledgements}
L.C.N.S. would like to thank Conselho Nacional de Desenvolvimento Cient\'ifico e Tecnol\'ogico (CNPq) for partial financial support through the research Project No. 164762/2020-5 and F.M.S. would like to thank CNPq for financial support through the research Project No. 165604/2020-4. I. P. L. was partially supported by the National Council for Scientific and Technological Development - CNPq grant 306414/2020-1 and by the grant 3197/2021, Para\'iba State Research Foundation (FAPESQ). I. P. L. would like to acknowledge the contribution of the COST Action CA18108. V.B.B. is partially supported by CNPq through the Research Project No. 307211/2020-7.

\bibliographystyle{ieeetr}
\bibliography{references.bib}

\begin{thebibliography}{10}

\bibitem{witt}
R.~Utiyama and B.~S. DeWitt, ``Renormalization of a classical gravitational
  field interacting with quantized matter fields,'' {\em Journal of
  Mathematical Physics}, vol.~3, no.~4, pp.~608--618, 1962.

\bibitem{fder1}
T.~P. Sotiriou and V.~Faraoni, ``{f(R) Theories Of Gravity},'' {\em Reviews of
  Modern Physics}, vol.~82, pp.~451--497, 2010.

\bibitem{vera}
V.~C. Rubin and W.~K. Ford~Jr, ``{Rotation of the Andromeda nebula from a
  spectroscopic survey of emission regions},'' {\em The Astrophysical Journal},
  vol.~159, p.~379, 1970.

\bibitem{cluster}
F.~Zwicky, ``Die rotverschiebung von extragalaktischen nebeln,'' {\em Helvetica
  Physica Acta}, vol.~6, pp.~110--127, 1933.

\bibitem{Perlmutter:1998np}
S.~Perlmutter, G.~Aldering, G.~Goldhaber, R.~A. Knop, P.~Nugent, P.~G. Castro,
  S.~Deustua, S.~Fabbro, A.~Goobar, D.~E. Groom, {\em et~al.}, ``{Measurements
  of $\Omega$ and $\Lambda$ from 42 high-redshift supernovae},'' {\em The
  Astrophysical Journal}, vol.~517, no.~2, p.~565, 1999.

\bibitem{Sahni:2004ai}
V.~Sahni, ``5 dark matter and dark energy,'' {\em The Physics of the Early
  Universe}, pp.~141--179, 2004.

\bibitem{capozziello2011extended}
S.~Capozziello and M.~De~Laurentis, ``Extended theories of gravity,'' {\em
  Physics Reports}, vol.~509, no.~4-5, pp.~167--321, 2011.

\bibitem{de2010f}
A.~De~Felice and S.~Tsujikawa, ``{$f(R)$ theories},'' {\em Living Reviews in
  Relativity}, vol.~13, no.~1, pp.~1--161, 2010.

\bibitem{nojiri2011unified}
S.~Nojiri and S.~D. Odintsov, ``{Unified cosmic history in modified gravity:
  from $F(R)$ theory to Lorentz non-invariant models},'' {\em Physics Reports},
  vol.~505, no.~2-4, pp.~59--144, 2011.

\bibitem{lobo2008dark}
F.~S.~N. Lobo, ``The dark side of gravity: Modified theories of gravity,'' {\em
  arXiv preprint arXiv:0807.1640}, 2008.

\bibitem{harada1998neutron}
T.~Harada, ``Neutron stars in scalar-tensor theories of gravity and catastrophe
  theory,'' {\em Physical Review D}, vol.~57, no.~8, p.~4802, 1998.

\bibitem{orellana2013structure}
M.~Orellana, F.~Garc{\'\i}a, F.~A.~T. Pannia, and G.~E. Romero, ``{Structure of
  neutron stars in $R$-squared gravity},'' {\em General Relativity and
  Gravitation}, vol.~45, no.~4, pp.~771--783, 2013.

\bibitem{momeni2015tolman}
D.~Momeni and R.~Myrzakulov, ``{Tolman–Oppenheimer–Volkoff equations in
  modified Gauss–Bonnet gravity},'' {\em International Journal of Geometric
  Methods in Modern Physics}, vol.~12, no.~02, p.~1550014, 2015.

\bibitem{oliveira2015neutron}
A.~Oliveira, H.~Velten, J.~Fabris, and L.~Casarini, ``Neutron stars in
  {R}astall gravity,'' {\em Physical Review D}, vol.~92, no.~4, p.~044020,
  2015.

\bibitem{hendi2016modified}
S.~Hendi, G.~Bordbar, B.~E. Panah, and S.~Panahiyan, ``Modified {TOV} in
  gravity's {R}ainbow: properties of neutron stars and dynamical stability
  conditions,'' {\em Journal of Cosmology and Astroparticle Physics},
  vol.~2016, no.~09, p.~013, 2016.

\bibitem{singh2019einstein}
K.~N. Singh, F.~Rahaman, and A.~Banerjee, ``{E}instein’s cluster mimicking
  compact star in the teleparallel equivalent of general relativity,'' {\em
  Physical Review D}, vol.~100, no.~8, p.~084023, 2019.

\bibitem{maurya2020charged}
S.~K. Maurya and F.~Tello-Ortiz, ``{Charged anisotropic compact star in
  $f(R,T)$ gravity: A minimal geometric deformation gravitational decoupling
  approach},'' {\em Physics of the Dark Universe}, vol.~27, p.~100442, 2020.

\bibitem{mota2019combined}
C.~E. Mota, L.~C.~N. Santos, G.~Grams, F.~M. da~Silva, and D.~P. Menezes,
  ``Combined {R}astall and {R}ainbow theories of gravity with applications to
  neutron stars,'' {\em Physical Review D}, vol.~100, no.~2, p.~024043, 2019.

\bibitem{paper1}
C.~E. Mota, L.~C.~N. Santos, F.~M. da~Silva, C.~V. Flores, T.~J.~N. da~Silva,
  and D.~P. Menezes, ``Anisotropic compact stars in {R}astall--{R}ainbow
  gravity,'' {\em Classical and Quantum Gravity}, vol.~39, no.~8, p.~085008,
  2022.

\bibitem{da2021rapidly}
F.~M. da~Silva, L.~C.~N. Santos, and C.~C. Barros, ``Rapidly rotating compact
  stars in {R}astall’s gravity,'' {\em Classical and Quantum Gravity},
  vol.~38, no.~16, p.~165011, 2021.

\bibitem{ns1}
A.~Cooney, S.~DeDeo, and D.~Psaltis, ``{Neutron stars in $f(R)$ gravity with
  perturbative constraints},'' {\em Physical Review D}, vol.~82, no.~6,
  p.~064033, 2010.

\bibitem{ns2}
S.~Capozziello, M.~De~Laurentis, R.~Farinelli, and S.~D. Odintsov,
  ``{Mass-radius relation for neutron stars in $f(R)$ gravity},'' {\em Physical
  Review D}, vol.~93, no.~2, p.~023501, 2016.

\bibitem{ns3}
S.~Arapo{\u{g}}lu, C.~Deliduman, and K.~Y. Ek{\c{s}}i, ``{Constraints on
  perturbative $f(R)$ gravity via neutron stars},'' {\em Journal of Cosmology
  and Astroparticle Physics}, vol.~2011, no.~07, p.~020, 2011.

\bibitem{ns4}
P.~H. R.~S. Moraes, J.~D.~V. Arba{\~n}il, and M.~Malheiro, ``{Stellar
  equilibrium configurations of compact stars in $f(R,T)$ theory of gravity},''
  {\em Journal of Cosmology and Astroparticle Physics}, vol.~2016, no.~06,
  p.~005, 2016.

\bibitem{ns5}
J.~M.~Z. Pretel, S.~E. Jor{\'a}s, R.~R.~R. Reis, and J.~D.~V. Arba{\~n}il,
  ``{Neutron stars in $f(R,T)$ gravity with conserved energy-momentum tensor:
  Hydrostatic equilibrium and asteroseismology},'' {\em Journal of Cosmology
  and Astroparticle Physics}, vol.~2021, no.~08, p.~055, 2021.

\bibitem{ns6}
S.~I. dos Santos, G.~A. Carvalho, P.~H. R.~S. Moraes, C.~H. Lenzi, and
  M.~Malheiro, ``{A conservative energy-momentum tensor in the $f(R,T)$ gravity
  and its implications for the phenomenology of neutron stars},'' {\em The
  European Physical Journal Plus}, vol.~134, no.~8, pp.~1--8, 2019.

\bibitem{ns7}
M.~Sharif and A.~Waseem, ``{Anisotropic quark stars in $f(R,T)$ gravity},''
  {\em The European Physical Journal C}, vol.~78, no.~10, pp.~1--10, 2018.

\bibitem{ns8}
D.~Deb, S.~V. Ketov, M.~Khlopov, and S.~Ray, ``{Study on charged strange stars
  in $f(R,T)$ gravity},'' {\em Journal of Cosmology and Astroparticle Physics},
  vol.~2019, no.~10, p.~070, 2019.

\bibitem{rastall}
P.~Rastall, ``{Generalization of the Einstein theory},'' {\em Physical Review
  D}, vol.~6, no.~12, p.~3357, 1972.

\bibitem{santos2022}
C.~E. Mota, L.~C.~N. Santos, F.~M. da~Silva, G.~Grams, I.~P. Lobo, and D.~P.
  Menezes, ``{Generalized Rastall's gravity and its effects on compact
  objects},'' {\em International Journal of Modern Physics D}, vol.~31, no.~04,
  p.~2250023, 2022.

\bibitem{Velten:2021xxw}
H.~Velten and T.~R.~P. Caram\^es, ``{To conserve, or not to conserve: A review
  of nonconservative theories of gravity},'' {\em Universe}, vol.~7, no.~2,
  p.~38, 2021.

\bibitem{harko2014f}
T.~Harko, F.~S.~N. Lobo, G.~Otalora, and E.~N. Saridakis, ``{$f(T,\CMcal{T})$
  gravity and cosmology},'' {\em Journal of Cosmology and Astroparticle
  Physics}, vol.~2014, no.~12, p.~021, 2014.

\bibitem{salako2020study}
I.~G. Salako, M.~Khlopov, S.~Ray, M.~Arouko, P.~Saha, and U.~Debnath, ``{Study
  on anisotropic strange stars in $f (T, T)$ gravity},'' {\em Universe},
  vol.~6, no.~10, p.~167, 2020.

\bibitem{Carroll2004}
S.~Carroll, {\em Spacetime and Geometry}.
\newblock Cambridge University Press, 2019.

\bibitem{PhysRevC.82.055803}
F.~J. Fattoyev, C.~J. Horowitz, J.~Piekarewicz, and G.~Shen, ``Relativistic
  effective interaction for nuclei, giant resonances, and neutron stars,'' {\em
  Physical Review C}, vol.~82, no.~5, p.~055803, 2010.

\bibitem{PhysRevC.99.045202}
O.~Louren{\c{c}}o, M.~Dutra, C.~H. Lenzi, C.~V. Flores, and D.~P. Menezes,
  ``{Consistent relativistic mean-field models constrained by GW170817},'' {\em
  Physical Review C}, vol.~99, no.~4, p.~045202, 2019.

\bibitem{PhysRevC.93.025806}
M.~Dutra, O.~Louren{\c{c}}o, and D.~P. Menezes, ``Stellar properties and
  nuclear matter constraints,'' {\em Physical Review C}, vol.~93, no.~2,
  p.~025806, 2016.

\bibitem{guichon}
P.~A.~M. Guichon, ``A possible quark mechanism for the saturation of nuclear
  matter,'' {\em Physics Letters B}, vol.~200, pp.~235--240, 1988.

\bibitem{saito94}
K.~Saito and A.~W. Thomas, ``{A quark-meson coupling model for nuclear and
  neutron matter},'' {\em Physics Letters B}, vol.~327, no.~1-2, pp.~9--16,
  1994.

\bibitem{saito95}
K.~Saito and A.~W. Thomas, ``{Composite nucleons in scalar and vector mean
  fields},'' {\em Physical Review C}, vol.~52, no.~5, p.~2789, 1995.

\bibitem{Pal95}
S.~Pal, M.~Hanauske, I.~Zakout, H.~St{\"o}cker, and W.~Greiner, ``{Neutron star
  properties in the quark-meson coupling model},'' {\em Physical Review C},
  vol.~60, no.~1, p.~015802, 1999.

\bibitem{Grams:2015inf}
G.~Grams, A.~M. Santos, and D.~P. Menezes, ``{Equation of State Grid with the
  Quark-Meson-Coupling Model},'' {\em Brazilian Journal of Physics}, vol.~46,
  no.~1, pp.~111--119, 2016.

\bibitem{bps}
G.~Baym, C.~Pethick, and P.~Sutherland, ``{The Ground state of matter at high
  densities: Equation of state and stellar models},'' {\em The Astrophysical
  Journal}, vol.~170, pp.~299--317, 1971.

\bibitem{Ozel_2016}
F.~\"Ozel and P.~Freire, ``{Masses, Radii, and the Equation of State of Neutron
  Stars},'' {\em Annual Review of Astronomy and Astrophysics}, vol.~54,
  pp.~401--440, 2016.

\bibitem{Steiner_2018}
A.~W. Steiner, C.~O. Heinke, S.~Bogdanov, C.~K. Li, W.~C. Ho, A.~Bahramian, and
  S.~Han, ``Constraining the mass and radius of neutron stars in globular
  clusters,'' {\em Monthly Notices of the Royal Astronomical Society},
  vol.~476, no.~1, pp.~421--435, 2018.

\bibitem{Cromartie}
H.~T. Cromartie, E.~Fonseca, S.~M. Ransom, P.~B. Demorest, Z.~Arzoumanian,
  H.~Blumer, P.~R. Brook, M.~E. DeCesar, T.~Dolch, J.~A. Ellis, {\em et~al.},
  ``{Relativistic Shapiro delay measurements of an extremely massive
  millisecond pulsar},'' {\em Nature Astronomy}, vol.~4, no.~1, pp.~72--76,
  2020.

\bibitem{PREX2021}
B.~T. Reed, F.~J. Fattoyev, C.~J. Horowitz, and J.~Piekarewicz, ``{Implications
  of PREX-2 on the equation of state of neutron-rich matter},'' {\em Physical
  Review Letters}, vol.~126, no.~17, p.~172503, 2021.

\bibitem{Demorest}
P.~B. Demorest, T.~Pennucci, S.~M. Ransom, M.~S.~E. Roberts, and J.~W.~T.
  Hessels, ``{A two-solar-mass neutron star measured using Shapiro delay},''
  {\em Nature}, vol.~467, no.~7319, pp.~1081--1083, 2010.

\bibitem{Antoniadis}
J.~Antoniadis, P.~C.~C. Freire, N.~Wex, T.~M. Tauris, R.~S. Lynch, M.~H.
  Van~Kerkwijk, M.~Kramer, C.~Bassa, V.~S. Dhillon, T.~Driebe, {\em et~al.},
  ``A massive pulsar in a compact relativistic binary,'' {\em Science},
  vol.~340, no.~6131, p.~1233232, 2013.

\end{thebibliography}

\end{document}